EDITORIAL

# Ten simple rules to create a serious game, illustrated with examples from structural biology

Marc Baaden[1], Olivier Delalande[2], Nicolas Ferey[3], Samuela Pasquali[4], Jérôme Waldispühl[5], Antoine Taly[1]*

**1** Laboratoire de Biochimie Théorique, Université Paris Diderot, Sorbonne Paris Cité, Paris, France, **2** Institut de Génétique et Développement de Rennes, Univ. Rennes 1, Rennes, France, **3** VENISE group, CNRS-LIMSI, Orsay, France, **4** Laboratoire de Cristallographie et RMN Biologiques, Faculté des sciences pharmaceutiques et biologiques, Université Paris Descartes et Université Sorbonne Paris Cité, Paris, France, **5** School of Computer Science, McGill University, Montréal, Canada

* taly@ibpc.fr

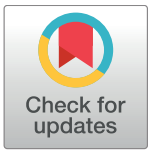

**Funding:** This work was supported by the "Initiative d'Excellence" program from the French State (Grant "DYNAMO", ANR-11-LABX-0011-01). The funders had no role in study design, data collection and analysis, decision to publish, or preparation of the manuscript.

**Competing interests:** The authors have declared that no competing interests exist.

## Overview

Serious scientific games are games whose purpose is not just fun. In the field of science, the serious goals include crucial activities for scientists: outreach, teaching and research. The number of serious games is increasing rapidly, in particular citizen science games (CSGs), games that allow people to produce and/or analyze scientific data. It is possible to build a set of rules providing a guideline to create or improve serious games. We present arguments gathered from our own experience (Phylo, DocMolecules, the HiRE-RNA contest and Pangu) as well as examples from the growing literature on scientific serious games.

## Introduction

Science has an enormous impact on society; therefore, understanding and participating in science projects is important for citizens. The involvement of nonscientists in the realization and design of science projects is called citizen science. Citizen science has become more abundant during the past decade [1]. One good example of a citizen science project is GalaxyZoo, which engages participants in classifying galaxies and has produced numerous publications—48 by 2014 [1].

This way of developing a research program is on the rise; one striking example is the game Foldit [2–3]. The Foldit project is an online 3D jigsaw puzzle in which players are invited to shake and wiggle the 3D structure of proteins to find their most stable conformations [4]. Since its release in 2008, the project has brought remarkable results from a biological point of view [2–3], but it was also useful to collaboratively develop new algorithms to solve a particular scientific problem [5]. Indeed, a very recent study showed that the results of players for model refinement tasks can be compared favorably with those of professional researchers [6]. Similarly, Mazzanti et al. developed the HiRE-RNA contest and showed that novice players are able to fold RNA structures without much prior knowledge [7].

The success of citizen science initiatives is at least partly related to the ability of groups to perform many tasks better than the sum of their individuals, the "wisdom of crowds effect." Many studies have gathered information on the determinants of collective intelligence [8], particularly thanks to controlled experiments in "crowd wisdom" [9]. These studies have shown that key variables need to be scrutinized, such as information network structure [10], communication between users [11], and social influence [12]. Based on these fundamental observations and our own experience, we present 10 simple rules to create or improve citizen science games (CSGs) in this developing field. We focus on the computational molecular biology area, in which CSGs are especially frequent. We share experience from our own Phylo, DocMolecules, HiRE-RNA, and Pangu projects and compare it to several other initiatives such as FoldIt and EteRNA. Such games may fall into multiple categories such as collecting scientific data, sorting it, or solving problems.

## Rule 1: Define a (serious) goal

The most synthetic definition of a serious game is that of video game designers Sande Chen and David Michael: a game "in which education (in its various forms) is the primary goal rather than entertainment" [13]. The work of Julian Alvarez, Damien Djaouti, and Olivier Rampnoux [14] further defined a serious game as a device, digital or otherwise, whose initial intention is to consistently combine utilitarian aspects with playful means. Such an association is aimed at an activity or a market other than entertainment alone. Therefore, a purpose needs to be clearly defined in terms of science, outreach, and teaching. A good game may address all three aspects. Knowing that professional video game production can cost millions of dollars, funding should also be taken into consideration from the very beginning of the project. The funding impacts all aspects of the project. Thus, having a clear idea helps designers to be realistic about the goals that can be achieved.

### Is it to produce scientific data?

Many CSGs have a simulation component that allows the players to interact with and/or produce scientific data. Therefore, CSGs should lead to discoveries that can ultimately be published in the scientific literature. The scientific relevance of the results of gameplay increases the player's interest and motivation.

We note that up to now, most of the publications on the various CSGs mainly concern the games themselves, discussing the quality of data generated, impact on motivation, etc. This outcome is expected, as most of the initial publications were proof of principle. So far, only a few games generated actual scientific results on the subject they were meant to study. One intriguing common point among the first projects that published data or results obtained using data generated by players on open-ended questions (e.g., Phylo, Foldit, Eyewire [15], and EteRNA) is that they all involved pattern-matching tasks. However, although serious games date back to before the 1980s, such games with a scientific twist are relatively young, such that conclusions are difficult to make.

### Is it outreach?

There is only a small step from a citizen science project to outreach, because the involvement of participants is a criterion for success in citizen science projects [16]. This natural link leads to adaptation opportunities. Outreach can be an objective per se, as in the case of DocMolecules, which uses simulation and visualization tools developed for other projects on interactive docking to convey the molecular-level action of a drug in the fight against allergies.

### Is it teaching?

Videogames have properties that make them adequate learning platforms [17]. Games developed originally for research are regularly used for teaching as well. Good examples are Foldit [18], Phylo [19], EteRNA [20], and the HiRE-RNA contest [7]. The difference between research and education use lies partially in the terms used, i.e., in-game actions are sometimes called by a name that is presumably more familiar to the players but that hides the correct scientific term. Unfortunately, this choice may have consequences for learning that can only be limited with a debriefing to make the right connection between game and course. Therefore, it could be advised not to sacrifice precision of scientific terms when players may use the game for educational purposes (as well as for people primarily interested in science). Another aspect to bear in mind is that the success of informal learning around games depends on players' profiles [21].

## Rule 2: Fine-tune the balance between entertainment and serious tasks

As mentioned above, a serious game is a chimera of a utilitarian goal and game mechanics. Ideally, the game design should be implemented as a function of the objectives of the game (e.g., data production, knowledge diffusion). Equilibrium and compromise need to be found between scientific accuracy and player accessibility ([22], p51). The tradeoff particularly applies to 1) visualization and graphics, 2) interaction design, and 3) scoring [4].

The level of simplification of scientific information is a key point. Not all players are looking for the same level of information. Therefore, providing access to more advanced material can help to keep experts around. For example, Phylo integrates an expert interface accessible to users who play at least 20 games with the classic edition, which allows them to play on larger grids (300 columns) than those used in the basic version (25 columns). This feature helps to increase the engagement of the most assiduous users.

Entertainment can also be used as a reward for achievements in the game. For example, short animated sequences related to the scientific topic can be both informative and entertaining at the end of a completed level in the game.

## Rule 3: Enable the player to interact with scientific data

Use of scientific data enables game designers to raise player interest and aim for participative data production of high scientific relevance. Intuitive interaction with the data (e.g., through molecular simulations) enriches the learning experience. One way to allow interaction with scientific data is to derive a "single quantitative metric of success" that facilitates the transformation of a task into a game element [23]. For instance, Phylo uses a simplified scoring function that estimates the quality of an alignment.

One route to this use of data is to recycle available simulation tools. The prototypical example is Foldit, which derives from the Rosetta@home program [24], which in turn builds on the Rosetta software [25]. Similarly, the software UnityMol [26] and BioSpring [27] were used to create DocMolecules. DocMolecules uses Protein Data Bank (PDB) structures as input for biological targets and molecular models for drugs. Use of force field terms such as nonbonded interactions and precomputed electrostatic fields drives feedback loops in the game.

When generating data, the considerations presented in "Ten Simple Rules for Effective Online Outreach" [28] apply, especially Rule #8 ("Collect and assess data"). An interesting aspect of CSGs, in which the data are generated by volunteers, is whether the data are made available. There is a general tendency to increase the availability of the data. For example, metrics measured in the game, and largely used for the development of the game itself, can be made open [29].

The availability of the data generated by the players paves the way for another change in perspective for the participants. In addition to receiving the data they generated predigested by professional scientists, players can also access untreated information that allows them to analyze these data. A striking example is the resolution of a previously unsolved crystal structure by scientists thanks to a Foldit-generated model [30]. Furthermore, the source code of some CSGs has been publicly released to foster the collective development of these platforms (e.g., Phylo, Mark2Cure).

## Rule 4: Promote onboarding and engagement

Players have heterogeneous expectations [31]. Therefore, the reward system should be versatile. The entry barrier should be low, and ideally, the difficulty is adapted to each player ([22], p49, 52). For example, the background of players (general public versus students) has been found to be related to players' abilities [32]. Defining player activities (gameplay) should build on tasks that the participants enjoy doing/completing ([22], p53).

It is often proposed that adding game components to a serious task should increase the motivation of users. However, it can also be argued that using this extrinsic motivation could interfere with the user's intrinsic motivation. In the case of CSGs, this could lead to alteration of the data generated by players. A recent study suggests that this concern is not necessarily justified [33].

It is crucial to provide feedback to the players about their progress. This can be done through competitive score production (e.g., best scores tables, online community), which may create a knock-on effect on players who will progress more quickly in knowledge and competence acquisition. Game mechanics can orient player contributions to allow covering a project need (as when Foldit encouraged solving many targets in the Critical Assessment of protein Structure Prediction [CASP] competition, a worldwide competition allowing laboratories to test their folding methods [34]). The overall aim is to increase player engagement. However, it should be kept in mind that player profiles favoring competition are only a fraction of the population. Nevertheless, for CSGs, as for other crowdsourcing programs, the results rely mainly on a few participants that contribute most of the data, called "whales" [35].

It should be kept in mind that the motivations of players are not those of professional scientists. For example, Foldit players offered the possibility to be coauthors of articles decided rather to be credited together with their Foldit team [23].

## Rule 5: Manage information flow

The information given to individuals in a group has an important impact on their collective behavior, i.e., the information to which they have access might push them to copy others, which could lead to specific behavior [11]. Exchange of information between participants, as well as between the system (the serious game and its backend) and participants, is a crucial point [23]. However, this dialogue can hardly be established solely by scores, which provide rather limited information and can even be misleading [7]. The exchange of data is part of what allows collective intelligence to emerge. Indeed, the ability to communicate has a significant impact on the emergence of collective intelligence [8], with the network structure having a strong influence [9]. This is illustrated by the study of Tinati et al., which noted positive effects of communication in web-based citizen science programs ([36], see also references cited therein). A good example is found in EteRNA, in which players up-vote their favorite solution.

Furthermore, specific behaviors were observed, allowing participants to be classified in categories: discoverer, hypothesizer, and investigator [37]. When a game allows discussions between members of the community, another category of participants mainly interested in using this media for socializing can also participate in the diffusion of ideas within the crowd. These behaviors can lead to collective intelligence. So far, CSGs with active forums tend to be those that produce articles on application results generated by participants.

Allowing communication between users can create bias and be at the origin of group thinking phenomena. This could be due to complementary effects such as social influence effects, the range reduction effect, and the confidence effect [12]. To avoid this, Amazon Mechanical Turk deliberately forbids communication between participants. On the other hand, allowing users to exchange information opens the door to more complex calculations [38].

## Rule 6: Provide an appropriate narrative

The narrative is an important aspect in many games [39]. This is also true for serious games: "Stories are equally important for serious and non-serious games alike" [40]. The plot should give sense and context to the game so that everything is connected coherently and the player knows his/her part in the overall story: "While we cannot always control the actions of the player or the way she plays the game, we can adjust our storytelling technique to better align our learning objectives with our dramatic objectives" [40]. A storyboard can prove helpful for development, particularly for games that must reconcile divergent serious and game objectives [41].

To better target the audience (depending on the project), it is necessary to convey the difficulty of progressing or switching between different levels. An introductory tutorial sequence helps to provide a progression in the dispensing of scientific information (both concepts and vocabulary). With DocMolecules, this is done through two separate levels of the game: the first level presents the context and cellular aspects, and the second level allows players to manipulate the drug molecule in order to dock it. However, the Foldit team has noticed that a tutorial may not be mandatory when the user can discover the rules through experimentation [42].

## Rule 7: Adapt your level design

Depending on the objective and the audience, the degree of simplification of scientific content and manipulation must be adjusted, which is essential to make the game accessible to a broad audience and maintain player interest [22].

Game duration needs to be adapted. The serious goal implies that the game needs to be adapted to everyone, including those who define themselves as non-gamers, which in turn implies to develop casual games. Casual games are easily accessed, simple to control, and not punishing [22].

In order to control the difficulty, ideally the gameplay should not exceed 5–20 minutes, and the game should be targeted to the right person, who will find it rewarding [22]. It is therefore important to be able to predict a task's difficulty, which is essential for channeling these tasks to players with the appropriate skill level [32].

## Rule 8: Develop good graphics, not just for eye candy

The primary scientific data is often complex; adapted graphical representations will help significantly to better understand it (Fig 1). High-quality graphics increase the player's immersion. In the context of displaying complex scientific objects such as molecules for example, shadowing is a necessary feature for volume rendering and shape perception. Adjusting the crowding of the game scene allows realistic rendering of molecular worlds [43]. Indeed, experience from scientific molecular movies indicates that there are several ways to represent biological molecules and their crowding [44].

In representing molecules, the game could switch between ball-and-stick and molecular surfaces depending on the size and nature of the molecules to be displayed (e.g., to distinguish small chemical compounds from large biological macromolecules). Hiding information

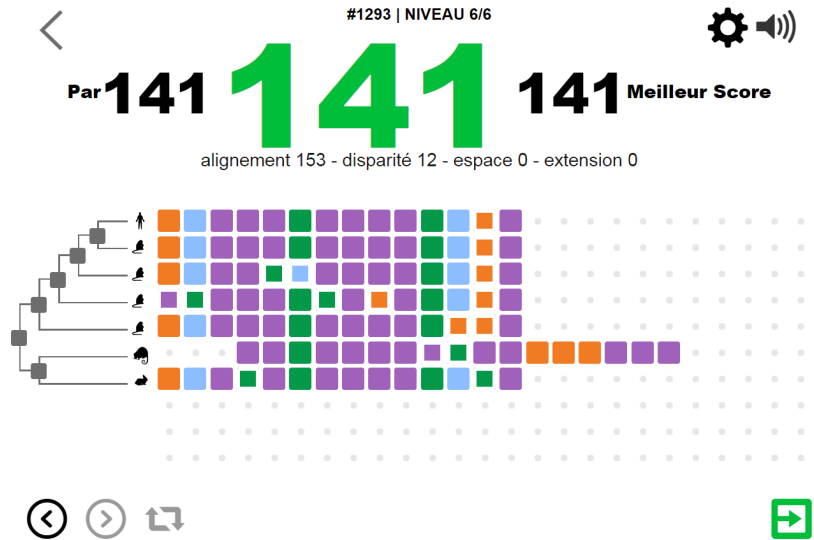

**Fig 1. Screenshot of Phylo illustrating the high level of information that can be provided to the player through graphics.**

https://doi.org/10.1371/journal.pcbi.1005955.g001

depending on player activities can also be used, as in the case of toggling on and off the side chain representation in Foldit [2].

## Rule 9: Use all modalities, particularly sound

The data a player will have to deal with in a serious game may be very complex and multidimensional [45]. If everything is conveyed visually, that channel may quickly become overloaded, and the player will be lost. One solution is to make use of several channels, not only the visual one, to convey important properties. Sound effects and music can drive interest or increase scenario effects. A good option is to simplify the rendering by conveying some information through sound rather than visual effects (e.g., score progress or formation of an interaction) [46].

Another modality that can be developed is touch (for example, through manipulation by hand with augmented reality applications) [47]. An example is presented in our Pangu project (Fig 2).

## Rule 10: "Iteratively assess what works and what doesn't"

To illustrate how important it is to prototype, evaluate, and iterate, we use here a title taken from the article "Ten Simple Rules for Effective Online Outreach" [28], in which this is Rule #9.

It is common practice to do iterations in game development [48], but in the case of serious games, this process involves three groups of evaluators instead of two. In addition to players and developers, scientists must come into play [4]. An interesting conclusion made by Cooper et al. [4] is that "we have also learned not to expect the way that expert scientists view the problem to be the best way for players." The iteration process can address many of the points described above, such as visualization and graphics, interaction design, and scoring mechanisms.

After creating a prototype, a critical step is to test it, which requires defining evaluation criteria for both game components and utilitarian aspects. The criteria defined to study the

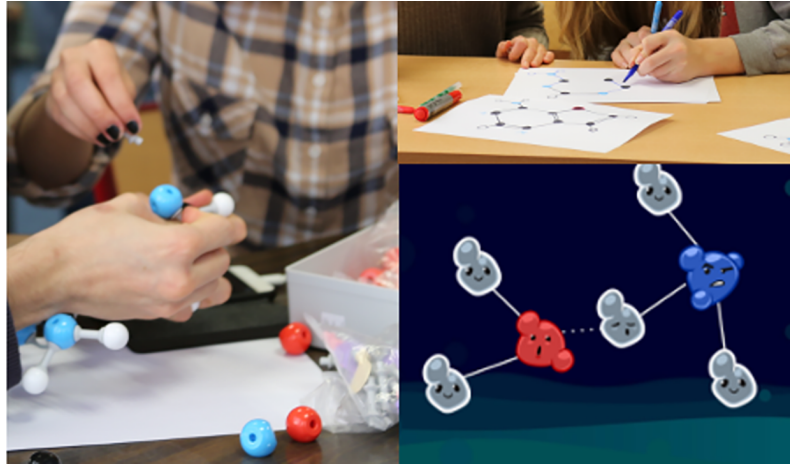

**Fig 2. Example of an augmented reality application (Pangu).** Players build their tangible model of the expected molecule with a standard molecular model kit (or draw it on paper) and scan it with a mobile device, which validates that the correct molecule was built.

https://doi.org/10.1371/journal.pcbi.1005955.g002

citizen science project Zooniverse, i.e., a "success matrix" measuring both contribution to science and public engagement, could be used as an example [16].

The final step is to iterate when appropriate. It seems natural that a CSG uses players to assess quality. Players may even become valuable contributors and propose original game content, as seen with Open-Phylo [49].

## Conclusion

The area of serious games has a long history, almost as long as video games themselves [50]. Recently, technological opportunities, including the internet, have allowed for the rapid expansion of CSGs [51]. This development represents a great opportunity by itself and should find more applications in the future with the democratization of virtual and augmented reality. Yet designing a good game remains a tricky business with many pitfalls. We hope that the guidelines provided above will help any scientific game designer to achieve successful implementation of a scientific endeavor within game mechanics.

## References

1. Follett R, & Strezov V (2015). An analysis of citizen science based research: usage and publication patterns. PLoS ONE, 10(11), e0143687. https://doi.org/10.1371/journal.pone.0143687 PMID: 26600041
2. Cooper S, Khatib F, Treuille A, Barbero J, Lee J, Beenen M, et al. (2010). Predicting protein structures with a multiplayer online game. Nature, 466(7307), 756–760. https://doi.org/10.1038/nature09304 PMID: 20686574
3. Good B M, & Su A I (2011). Games with a scientific purpose. Genome biology, 12(12), 135. https://doi.org/10.1186/gb-2011-12-12-135 PMID: 22204700
4. Cooper, S, Treuille, A, Barbero, J, Leaver-Fay, A, Tuite, K, Khatib, F., et al. The challenge of designing scientific discovery games. In Proceedings of the Fifth international Conference on the Foundations of Digital Games (pp. 40–47). ACM.
5. Khatib F, Cooper S, Tyka M D, Xu K, Makedon I, Popović Z, et al. (2011). Algorithm discovery by protein folding game players. Proceedings of the National Academy of Sciences, 108(47), 18949–18953.
6. Horowitz S, Koepnick B, Martin R, Tymieniecki A, Winburn AA, Cooper S, et al., University of Michigan students, Popović Z., Baker D., Khatib F., & Bardwell J.C.A. (2016). Determining crystal structures through crowdsourcing and coursework. Nature communications, 7, 12549. https://doi.org/10.1038/ncomms12549 PMID: 27633552

7. Mazzanti L, Doutreligne S, Gageat C, Derreumaux P, Taly A & Baaden M. What can human-guided simulations bring to RNA folding? Biophysical Journal 113 (2) 302–312. https://doi.org/10.1016/j.bpj.2017.05.047 PMID: 28648754

8. Woolley AW, Chabris CF, Pentland A, Hashmi N, & Malone TW (2010). Evidence for a collective intelligence factor in the performance of human groups. Science, 330(6004), 686–688. https://doi.org/10.1126/science.1193147 PMID: 20929725

9. Centola D, & Baronchelli A (2015). The spontaneous emergence of conventions: An experimental study of cultural evolution. Proceedings of the National Academy of Sciences, 112(7), 1989–1994.

10. Evans JA, & Foster JG (2011). Metaknowledge. Science, 331(6018), 721–725. https://doi.org/10.1126/science.1201765 PMID: 21311014

11. Laland KN (2004). Social learning strategies. Learning & behavior, 32(1), 4–14.

12. Lorenz J, Rauhut H, Schweitzer F, & Helbing D (2011). How social influence can undermine the wisdom of crowd effect. Proceedings of the National Academy of Sciences, 108(22), 9020–9025.

13. Michael D, Chen S (2005), Serious games that educate train and inform, Course Technology PTR, USA

14. Alvarez J, Djaouti D, & Rampnoux O (2016). Apprendre avec les serious games? Poitiers: Canopé Edition.

15. Kim JS, Greene MJ, Zlateski A, Lee K, Richardson M, Turaga SC, et al. (2014). Space-time wiring specificity supports direction selectivity in the retina. Nature, 509(7500), 331–336. https://doi.org/10.1038/nature13240 PMID: 24805243

16. Graham GG, Cox J, Simmons B, Lintott C, Masters K, Greenhill A, et al. (2015) How is success defined and measured in online citizen science: a case study of Zooniverse projects. Computing in science and engineering, PP (99) (22). ISSN 1521-9615 https://doi.org/10.1109/MCSE.2010.27

17. Gee JP (2003). What video games have to teach us about learning and literacy. Computers in Entertainment (CIE), 1(1), 20–20.

18. Franco J (2012). Online gaming for understanding folding, interactions, and structure. Journal of Chemical Education, 89(12), 1543–1546.

19. Kawrykow A, Roumanis G, Kam A, Kwak D, Leung C, Wu C, et al. (2012). Phylo: a citizen science approach for improving multiple sequence alignment. PLoS ONE, 7(3), e31362. https://doi.org/10.1371/journal.pone.0031362 PMID: 22412834

20. Lee J, Kladwang W, Lee M, Cantu D, Azizyan M, Kim H, et al. (2014). RNA design rules from a massive open laboratory. Proceedings of the National Academy of Sciences, 111(6), 2122–2127.

21. Iacovides I, McAndrew P, Scanlon E, & Aczel J (2014). The gaming involvement and informal learning framework. Simulation & Gaming, 45(4–5), 611–626.

22. Law E, & Ahn LV (2011). Human computation. Synthesis Lectures on Artificial Intelligence and Machine Learning, 5(3), 1–121.

23. Cooper S, Khatib F, & Baker D (2013). Increasing public involvement in structural biology. Structure, 21 (9), 1482–1484. https://doi.org/10.1016/j.str.2013.08.009 PMID: 24010706

24. Das R, Qian B, Raman S, Vernon R, Thompson J, Bradley P.,et al.(2007). Structure prediction for CASP7 targets using extensive all-atom refinement with Rosetta@ home. Proteins: Structure, Function, and Bioinformatics, 69(S8), 118–128.

25. Das R, & Baker D (2008). Macromolecular modeling with rosetta. Annu. Rev. Biochem., 77, 363–382. https://doi.org/10.1146/annurev.biochem.77.062906.171838 PMID: 18410248

26. Doutreligne S, Gageat C, Cragnolini T, Taly A, Pasquali S, Baaden M, et al. UnityMol: Simulation et Visualisation Interactive à des fins d'Enseignement et de Recherche. GGMM 2015, 118.

27. Ferey N, Delalande O, & Baaden M (2012). Biospring: an interactive and multi-resolution software for flexible docking and for mechanical exploration of large biomolecular assemblies. JOBIM 2012-Journées Ouvertes en Biologie, Informatique et Mathématiques, 433–434.

28. Bik HM, Dove ADM, Goldstein MC, Helm RR, MacPherson R, Martini K, et al. (2015) Ten Simple Rules for Effective Online Outreach. PLoS Comput Biol 11(4): e1003906. https://doi.org/10.1371/journal.pcbi.1003906 PMID: 25879439

29. Himmelstein J, Goujet R, Duong TK, Bland J, Lindner AB. Human Computation (2016) 3:1:119–141 2016.

30. Gilski M, Kazmierczyk M, Krzywda S, Zábranská H, Cooper S, Popović Z, et al. (2011). High-resolution structure of a retroviral protease folded as a monomer. Acta Crystallographica Section D: Biological Crystallography, 67(11), 907–914.

31. Bartle, RA (1990). Who Plays MUAs? Comms Plus!, October/November 1990 18–19.

**32.** ] Singh, Ahsan, Blanchette, Waldispühl (To Appear). Lessons from an online massive genomics computer game. In Proceedings of the 5th AAAI Conference on Human Computation and Crowdsourcing (HCOMP 2017).

**33.** Prestopnik N, Crowston Kevin, Wang Jun, Gamers, citizen scientists, and data: Exploring participant contributions in two games with a purpose, Computers in Human Behavior, Volume 68, March 2017, Pages 254–268.

**34.** Moult J, Fidelis K, Kryshtafovych A, Schwede T, & Tramontano A (2017). Critical assessment of methods of protein structure prediction (CASP)—Round XII. *Proteins*: *Structure*, *Function*, *and Bioinformatics*.

**35.** Sauermann H, & Franzoni C (2015). Crowd science user contribution patterns and their implications. Proceedings of the National Academy of Sciences, 112(3), 679–684.

**36.** Tinati, R, Simperl, E, & Luczak-Roesch, M (2017). To help or hinder: Real-time chat in citizen science.

**37.** Tinati, R, Simperl, E, Luczak-Roesch, M, Van Kleek, M, & Shadbolt, N (2014). Collective Intelligence in Citizen Science—A Study of Performers and Talkers. arXiv preprint arXiv:1406.7551.

**38.** Tremblay-Savard, O, Butyaev, A and J Waldispühl. 2016. Collaborative Solving in a Human Computing Game Using a Market, Skills and Challenges. In Proceedings of the 2016 Annual Symposium on Computer-Human Interaction in Play (CHI PLAY '16). ACM, New York, NY, USA, 130–141

**39.** Jenkins H (2004). Game design as narrative architecture. Computer, 44, 53

**40.** McDaniel R, Fiore SM, & Nicholson D (2010). Serious storytelling: Narrative considerations for serious games researchers and developers. In Serious game design and development: Technologies for training and learning (pp. 13–30). IGI Global.

**41.** Belanich J, Orvis KB, Horn DB, & Solberg JL (2011). Bridging game development and instructional design. In Instructional Design: Concepts, Methodologies, Tools and Applications (pp. 464–479). IGI Global.

**42.** Andersen, E, O'Rourke, E, Liu, YE, Snider, R, Lowdermilk, J, Truong, D, et al. (2012, May). The impact of tutorials on games of varying complexity. In Proceedings of the SIGCHI Conference on Human Factors in Computing Systems (pp. 59–68). ACM.

**43.** Ellis RJ (2001). Macromolecular crowding: obvious but underappreciated. Trends in biochemical sciences, 26 (10), 597–604. PMID: 11590012

**44.** McGill G (2008). Molecular movies. . . coming to a lecture near you. Cell, 133(7), 1127–1132. https://doi.org/10.1016/j.cell.2008.06.013 PMID: 18585343

**45.** O'Donoghue SI, Gavin AC, Gehlenborg N, Goodsell DS, Hériché JK, Nielsen CB, et al. (2010). Visualizing biological data—now and in the future. Nature methods, 7, S2–S4. https://doi.org/10.1038/nmeth.f.301 PMID: 20195254

**46.** Ballweg, H, Bronowska, AK, & Vickers, P (2017). Interactive Sonification for Structural Biology and Structre-based Drug Design.

**47.** Férey N, Nelson J, Martin C, Picinali L, Bouyer G, Tek A, et al. (2009). Multisensory VR interaction for protein-docking in the CoRSAIRe project. Virtual Reality, 13(4), 273.

**48.** O'Hagan AO, & O'Connor RV (2015). Towards an Understanding of Game Software Development Processes: A Case Study. In European Conference on Software Process Improvement (pp. 3–16). Springer International Publishing.

**49.** Kwak D, Kam A, Becerra D, Zhou Q, Hops A, Zarour E, et al. (2013). Open-Phylo: a customizable crowd-computing platform for multiple sequence alignment. Genome biology, 14(10), R116. https://doi.org/10.1186/gb-2013-14-10-r116 PMID: 24148814

**50.** Djaouti D, Alvarez J, Jessel JP, & Rampnoux O (2011). Origins of serious games. In Serious games and edutainment applications (pp. 25–43). Springer London.

**51.** Curtis V (2014). Online citizen science games: opportunities for the biological sciences. Applied & translational genomics, 3(4), 90–94.